\documentclass[11pt,a4paper]{JHEP3}

\usepackage[latin1]{inputenc}
\usepackage[english]{babel}
\usepackage{amsmath,amssymb}
\usepackage{graphicx}
\usepackage{cite}

\newcommand{\as}{\alpha_s}

\newcommand{\beq}{\begin{equation}}
\newcommand{\eeq}{\end{equation}}
\newcommand{\bea}{\begin{eqnarray}}
\newcommand{\eea}{\end{eqnarray}}
\newcommand{\bdm}{\begin{displaymath}}
\newcommand{\edm}{\end{displaymath}}
\def\as{\alpha_s}

\def\ord{{\cal O}}

\def\d{\partial}

\def \d{{\rm d} }
\def \d0 {D\O \;}

\title{Probing the low transverse momentum domain of $Z$ production with novel variables}

\author{Andrea Banfi$^a$, Mrinal Dasgupta$^b$, Simone Marzani$^{c}$ and Lee Tomlinson$^b$ \\

$^a$Institute for Theoretical Physics, ETH Zurich,  8093 Zurich, Switzerland\\
$^b$School of Physics \& Astronomy, University of Manchester, Manchester, M13 9PL, UK
 \\$^c$Institute for Particle Physics Phenomenology, Durham University, Durham DH1 3LE, UK
 \\ \\
 \email{banfi@itp.phys.ethz.ch \\mrinal.dasgupta@manchester.ac.uk \\simone.marzani@durham.ac.uk \\ lee.tomlinson@hep.manchester.ac.uk }}

\preprint{DCPT/11/122 \\ IPPP/11/61 \\ MAN/HEP/2011/15}

\keywords{QCD, NLO Computations, Hadronic Colliders, Standard Model}

\abstract{
The measurement of the low transverse momentum region of vector boson production in Drell--Yan processes has long been invaluable to testing our knowledge of QCD dynamics both beyond fixed-order in perturbation theory as well as in the 
non-perturbative region. Recently the D\O\ collaboration have introduced novel variables which lead to improved measurements compared to the case of the standard $Q_T$ variable. To complement this improvement on the experimental side, we develop here a complete phenomenological study dedicated in particular to the new $\phi^*$ variable. 
We compare our study, which contains the state-of-the-art next-to--next-to-leading resummation of large logarithms and a smooth matching to the full next-to-leading order $\left (\mathcal{O}\left( \alpha_s^2 \right) \right)$ result, to the experimental data and find excellent agreement over essentially the entire range of 
$\phi^*$, even without direct inclusion of non-perturbative effects. We comment on our findings and on the potential for future studies to constrain non-perturbative behaviour. 
}

\begin{document}

%======================================================================
\section{Introduction}
\label{sec:introduction}
The transverse momentum ($Q_T$) distribution of lepton pairs, or equivalently vector bosons, produced via the Drell-Yan process \cite{DY} has been a classic observable in the realm of phenomenological studies of QCD at hadron colliders, 
both in and beyond the perturbative domain with pioneering studies commencing over three decades ago \cite{DDT,APP,PP,Davies,DSW,CSS}. In spite of being well-studied, the Drell-Yan $Q_T$ variable remains of continuing importance 
both on the QCD theory side as well as for high precision Standard Model phenomenology, such as $W$ mass determination, and hence with important implications for Higgs studies as well as potential studies involving new particles that may be discovered for instance at the LHC.

As far as QCD theory is concerned, while the high $Q_T$ tail of the distribution ought to be described within fixed-order perturbative methods, the low $Q_T$ region is enriched by the presence of large logarithms which require resummation in order to yield a meaningful result. A successful description of the entire $Q_T$ distribution hence requires resummation (at least to next-to-leading (NLL) accuracy) accompanied by matching to, ideally, NLO fixed-order results.  In fact the state of the art for the $Q_T$ variable as far as resummation is concerned is up to the NNLL level \cite{Davies,CatGrazll,BecNeu} and fixed-order codes such as MCFM \cite{mcfm}, DYNNLO \cite{dynnlo} and FEWZ \cite{FEWZ} offer the required full NLO results for the differential distribution. Thus, on the theoretical side, the level of accuracy one can achieve in the study of the $Q_T$ variable is paralleled by few other observables in QCD. Successful confrontation of such calculations with data indicates control over QCD dynamics at the high precision level and paves the way for extending such studies to related 
observables at hadron colliders, such as the Higgs $Q_T$ distribution \cite{CataniHiggs}. There are also relatively new calculational approaches such as those of soft-collinear effective theory where the computation of the $Q_T$ spectrum has been carried out \cite{BecNeu,BecNeu2,pet} and yields comparable results to traditional resummation approaches as well as yielding the full calculation of the next-to--next-to-leading logarithms, a piece of which had not previously been computed till date\cite{BecNeu}.

Armed with such accurate perturbative predictions one should be ideally placed to explore the role of non-perturbative (NP) effects and either extract them from the data or at the very least set limits on their size. One may visualise 
the NP behaviour as, for instance, an intrinsic Fermi motion of partons within the proton which leads to a Gaussian smearing of the perturbative $Q_T$ spectrum, with comparisons to data offering an opportunity to constrain the parameters of the Gaussian. A commonly used parametrisation of non-perturbative effects is 
the Brock--Landry--Nadolsky--Yuan (BLNY) form factor \cite{BLNY} which was obtained by comparing the resummation formalism of Ref.~\cite{CSS}, as encoded in the RESBOS generator \cite{resbos1,resbos2}, to Tevatron Run-I data.

Alternatively, one can take the perturbative calculation as a means of directly approaching NP behaviour and study the ambiguities inherent in the perturbative approach (due to the divergence of the perturbative coupling). For examples of such studies we point the reader to Ref.~\cite{SmyGuff} for the $Q_T$ case and the detailed exposition in Ref.~\cite{DokWebMar} for the related case of energy-energy correlation in $e^{+}e^{-}$ annihilation as well as references therein. 

It is perhaps worth noting here that various approaches to $Q_T$ resummation 
do not concur on the size of non-perturbative effects required to describe experimental data. As an instance of this we note that in Ref.~\cite{FlorenceDY} an essentially perturbative approach was seen to give an adequate description of Tevatron Run-II data except at the very lowest $Q_T$ values where one may expect very strong NP effects. As a general remark, in the above context, before reaching firm conclusions on the size of NP effects (which one should in any case expect to vary from observable to observable and depend on the process) one should be sure about the robustness of the perturbative result and the uncertainty associated to it, which can be ascertained by varying the various perturbative scales in the problem. We shall comment in more detail on the role of these scale variations later in this article. Lastly, we should also point out that given the sensitivity of the $Q_T$ spectrum to QCD dynamics beyond fixed-order, it is an ideal variable for the testing of Monte Carlo event generators and for tuning the parameters therein as well as for testing new models of non-perturbative behaviour \cite{SeySiod} encoded in QCD event generators.

Recently, the D\O\ collaboration have introduced two new variables $a_T$ and $\phi^{*}$ which have been shown to offer advantages over the standard $Q_T$ variable in that they can be measured more accurately and hence offer the potential to push the theoretical studies further, yielding valuable new information on both perturbative and non-perturbative effects \cite{WV,WVBRW,D0phi}. As a concrete example of the utility and discriminating power of the new variables one can consider the issue of the small-$x$ broadening of non-perturbative effects suggested initially in comparisons of the RESBOS generator to semi-inclusive DIS data \cite{QTsmallx}. Since such small-$x$ effects could have a serious impact on the vector boson and Higgs resummed $Q_T$ spectra, especially in forward (high rapidity) regions and, in particular, at the LHC, it is clearly of importance to ascertain their presence or otherwise. Based on comparisons to their new data on $\phi^*$ the D\O\ collaboration were able to demonstrate that such effects were in fact disfavoured, which had not been previously possible due to errors on the $Q_T$ spectrum even with Tevatron Run-II data \cite{D0phi}.

In previous papers \cite{BDDaT,BDMphi} we have provided the details of a resummed treatment of the new variables and discussed their relationship to $Q_T$ and to each other. We have computed the resummation to NNLL accuracy and carried out the matching to fixed-order NLO results from MCFM. Since the variables $a_T/M$ and $\phi^*$ are essentially identical at low $Q_T$ 
\cite{WVBRW,D0phi,BDMphi} we shall focus here on the $\phi^*$ case which is also the variable favoured by the D\O\ collaboration in terms of measurement\cite{D0phi}. In the present paper we extend our initial theoretical studies to the phenomenological level by deriving our matched resummed results differentially in the vector boson rapidity and with precisely the same cuts as adopted for the D\O\ measurements \cite{D0phi}. We study the role of various scales in the problem including renormalisation and factorisation scale variation as well as resummation scale uncertainties (rescaling the argument of the logarithms we are resumming) and derive uncertainty bands for the perturbative result. We then compare our 
results to the experimental data for both electrons and muons in various rapidity bins. 

We organise the present paper as follows: in the next section we provide a reminder of the details of the observables and their dependence on soft emissions 
as well as write down the resummed formula we derived in our previous work. In the following section we consider the general full NNLL result with variations of factorisation, renormalisation and resummation scales so as to derive the uncertainty on the perturbative result. Next we provide our 
comparisons to the experimental data and comment on the quality of the agreement as well as the potential need to include non-perturbative effects before providing a concluding discussion identifying future developments. We also provide for convenience an appendix where we list the main formulae we use in this work. 

\section{The $\phi^*$ variable and its resummation}
In this section we remind the reader of the main features of the resummed result computed in Ref.~\cite{BDMphi}. There we derived the dependence of the $a_T$ and $\phi^*$ variables on soft and, optionally, collinear gluon emissions from the incoming partons. In both cases one found that, in the soft limit, the dependence on emissions was essentially via a single component of gluon transverse momentum which was the one normal to the axis defined by the nearly back-to-back leptons. Here, and for the rest of this paper, we shall focus on $\phi^*$, which is a measure of the deviation of the lepton opening angle $\Delta \phi$ in the transverse plane from its value at Born level, $\Delta \phi = \pi$. The $\phi^*$ variable is defined as
\begin{equation}
\phi^* = \tan \left (\phi_{\mathrm{acop}}/2 \right) \sin \theta^* = \left | \sum_i \frac{k_{Ti}}{M} \sin \phi_i \right | +\ord\left(\frac{k_{Ti}^2}{M^2}\right)\,,
\end{equation}
where $\phi_{\mathrm{acop}} =\pi -\Delta \phi$ is the acoplanarity angle which vanishes at Born level, $\sin \theta^*$ derives from the angle in a boosted frame such that the leptons make angles $\theta^*$ and $\pi-\theta^*$ with the beam\footnote{Our definition of $\phi^*$ corresponds to the variable $\phi^*_\eta$ in the \d0 study.}, $M$ is the mass of the lepton pair and $\phi_i$ is the angle of the gluon $i$ with respect to the lepton axis in the transverse plane, with $k_{Ti}$ the magnitude of its transverse momentum with respect to the emitting (incoming) partons. Requiring $\phi^*$ to be equal to some fixed value thus involves the constraint 
\begin{equation}
\label{eq:oned}
\delta\left( \phi^*-\frac{\left |\sum_i k_{yi} \right |}{M}\right) = \frac{1}{\pi} \int_{-\infty}^{\infty}
{\rm d} b\,M \cos(b M \phi^*) \prod_i e^{i b k_{yi}} \,,
\end{equation}
where on the RHS we note that the sum over gluon emissions appears in 
factorised form in impact parameter ($b$) space as a product over emissions. When combined with the factorisation of the matrix element squared in the soft-collinear limit one can exponentiate the single-emission contribution (with account of running coupling) in $b$ space. This results in the resummed form for the $\phi^*$ distribution which reads \cite{BDMphi}
\begin{equation}
\label{eq:resummed}
\frac{ {\rm d} \sigma}{ {\rm d} \phi^*} \left( \phi^*,M, \cos \theta^*, y \right) = \frac{\pi \alpha^2}{s N_c} 
\int_0^{\infty} {\rm d} b\, M \,\cos \left(bM \phi^* \right) 
e^{-R(\bar{b},M)} \nonumber \\ \times\,\Sigma \left(x_1,x_2,\cos\theta^*, b,M\right)\,,  
\end{equation}
where
\begin{equation}
x_{1,2} = \frac{M}{\sqrt{s}}e^{\pm y} \quad {\rm and} \quad \bar{b}= \frac{b e^{\gamma_E}}{2}\,.
\end{equation}
The expression above is yet to be integrated over the dilepton invariant mass $M$, the scattering angle $\theta^*$ and rapidity of the dilepton system (or equivalently the $Z$ boson rapidity) $y$. In our previous work we took into account the experimental cuts over $M$ and $\theta^*$ but we integrated our expression over the full rapidity range. In this paper we will present the results in different rapidity bins, to able to compare to the data.  

The function $R(b)$ in the exponent in  Eq.~(\ref{eq:resummed}) represents the single gluon emission contribution taking into account the running of $\alpha_s$. To be precise one can write
\begin{equation} \label{radiator}
R(b) = L g^{(1)}(\alpha_s L) +g^{(2)} (\alpha_s L) +\frac{\alpha_s}{\pi} g^{(3)} (\alpha_s L)+\cdots
\end{equation}
where $L= \ln\left(\bar{b}^2 M^2\right)$.
The functions $L g^{(1)}$, $g^{(2)}$ and $\frac{\alpha_s}{\pi} g^{(3)}$ are the leading, next-to-leading and 
next-to--next-to-leading logarithmic contributions respectively. Further sub-leading terms are not shown, as they are beyond our current accuracy. The quantity $\Sigma$ is basically the Born level result with the modification that the argument of the parton distribution functions (pdfs) entering therein is set as $1/b^2$ rather than a scale of the order of $M^2$, the dilepton invariant mass squared. This modification of pdfs is due to DGLAP resummation of logarithms arising from hard collinear emission from the incoming legs and coincides with the standard treatment for the $Q_T$ variable.  Also included in $\Sigma$ are leading-order (${\mathcal{O}}\left(\alpha_s \right)$) coefficient functions that are convolved into the pdfs. The precise form of $\Sigma$ can be found in Ref.~\cite{BDMphi} and in Appendix~\ref{app:res} for convenience.
The radiator $R(b)$ turns out to be precisely the same as for the $Q_T$ variable \cite{BDMphi} and hence the only difference from the $Q_T$ case is the presence of the cosine function here rather than the standard Bessel function.

In the current paper we present numerical results including full NNLL accuracy for $R(b)$, whereas in our previous work we had kept only the leading term in $\alpha_s$ of the NNLL function $g^{(3)} \sim \alpha_s^2 L$. The reason we have now used the full form of $g^{(3)}$ is that, due to recent work \cite{BecNeu} for the $Q_T$ variable, the complete result for $g^{(3)}$ has become available, whereas previous results had assumed (in the absence of an appropriate calculation) that one of the coefficients involved in $g^{(3)}$, which pertains to terms starting at order $\alpha_s^3$, would be the same as for the resummation of threshold logarithms. This turns out not to be the case and hence we use the recently computed result, which however does not have a visible numerical impact on our final results. 

Finally, an important issue that we handle in the next section is that thus far we have set all ambiguous scales in the problem to be of the order of the hard scale of the process, $M$. In reality our result is arbitrary beyond the NNLL 
accuracy of our resummation and the NLO accuracy of our fixed-order result in the standard way for any truncated fixed-order estimate. To assess 
this uncertainty is of course important before addressing the data and drawing conclusions, for instance, on the role of higher-order and sub-leading logarithmic terms omitted from our treatment and certainly before making any conclusive statement on non-perturbative effects. In the following section we thus also consider the role of scale variations on our resummed results for factorisation and 
renormalisation scales as well as varying the argument of the logarithm by a factor in a way that one generates sub-leading terms beyond NNLL, whose size can therefore be naturally estimated.

\section{Numerical evaluation, fixed-order matching and results}
The function $R(b)$ can  be found in Appendix~\ref{app:radiator} where we list the results for $g^{(1)}$, $g^{(2)}$ and $g^{(3)}$ as a function of the single logarithmic variable $\lambda = \beta_0 \alpha_s (M) \ln \left(  \bar{b}^2 M^2 \right)$. They are meaningful only in the range $0< \lambda < 1$, with the upper limit corresponding to the position of a divergence associated to the Landau pole, where at the one-loop level $\bar{b} = {\Lambda_{\mathrm{QCD}}}^{-1}$. The lower limit at which $R(b)$ vanishes is instead outside 
the jurisdiction of resummation. To handle these difficulties we need to limit the range of the $b$ integral (see for instance Ref.~\cite{DokWebMar}), such that $b_{\mathrm{min}}<b<b_{\mathrm{max}}$ with $b_{\mathrm{min}}= 2 M^{-1}e^{-\gamma_E}$ and $b_{\mathrm{max}}=  2 M^{-1}e^{-\gamma_E} \exp \left({1/(2 \beta_0 \alpha_s)}\right)$. 
For our our current work, in the region below $b=b_{\min}$ we simply set the radiator to zero while for $b>b_{\mathrm{max}}$ one has no contribution since we assume the radiator to be infinite here. The precise choice we make for $b_{\mathrm{max}}$  corresponds to a cut-off on perturbative dynamics in the vicinity of the Landau pole and we note that in practice increasing $\bar{b}_{\mathrm{max}}$ beyond $1/(3 \Lambda_{\mathrm{QCD}})$, the value we adopt here,  has a negligible impact on our result. 

In principle of course this is not a rigorously justified procedure and one is free to choose other prescriptions to regulate the divergence of the perturbative result and replace it with genuine non-perturbative effects. In practice one may view the dependence on the precise position of $b_{\mathrm{max}}$ as a sign of the sensitivity to non-perturbative dynamics and in our particular case varying the position of $b_{\mathrm{max}}$ in the vicinity of the Landau pole does not change our numerical results significantly. 
Also related to this issue is the evolution of the pdfs down to the scale $1/b$. In principle, we should evolve the pdfs down to $1/b_{ \rm max}$. In practice, we freeze the pdfs at a scale of the order $1$~GeV, which is the minimum value for which the pdfs of the set we consider are tabulated. However, by evolving the pdfs with the code HOPPET~\cite{hoppet}, we have checked that pushing the freezing scale below this value has negligible effects. We shall comment more on non-perturbative effects in the concluding section of this article.

To obtain results which can be compared to the data over a wide range of $\phi^*$, one needs to combine the resummed results with those from fixed-order codes which compute the $\phi^*$ distribution up to order $\alpha_s^2$, i.e. at the NLO level. In this context we note that our NNLL resummed result guarantees control over all logarithms that can arise up to NLO accuracy including the least singular $\alpha_s^2 L$ term. Having full control over all the logarithms at the two-loop order allows for a very simple matching formula 
\begin{equation}
\left(\frac{ {\rm d} \sigma}{{\rm d} \phi^*}\right)_{\mathrm{matched}} = \left(\frac{{\rm d} \sigma}{ {\rm d} \phi^*}\right)_{\mathrm{resummed }} +\left(\frac{{\rm d} \sigma}{{\rm d} \phi^*}\right)_{\mathrm{fixed \; order}}-\left(\frac{{\rm d} \sigma}{{\rm d} \phi^*}\right)_{\mathrm{expanded}},
\end{equation}
where one simply combines the resummation with the fixed-order and subtracts 
the expansion of the resummation to order $\alpha_s^2$ to remove any double counting. We note that for the matching to be considered successful at small $\phi^*$ the expansion of the resummation and the fixed-order result should cancel and the matched result should thus follow the pure resummed curve while at large $\phi^*$ the resummation should largely cancel against its expansion 
(up to relatively small order $\alpha_s^3$ terms) and the result should tend to the pure fixed-order curve.

\begin{figure}
\begin{center}
\includegraphics[width=0.7 \textwidth]{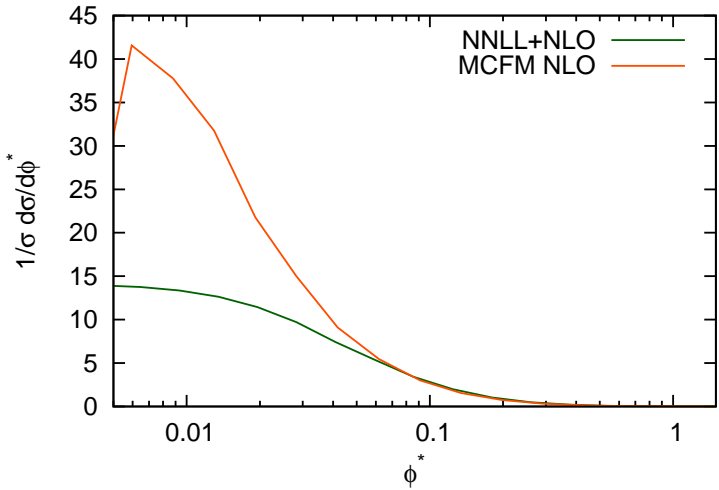}
\caption{Figure illustrating the full NNLL resummed result to NLO for the $\phi^*$ distribution. Also shown for comparison is the fixed-order result from MCFM while the range of $\phi^*$ chosen is that over which data has been collected \cite{D0phi}.}  
\label{fig:resumnlo}
\end{center}
\end{figure}

Fig.~\ref{fig:resumnlo} shows our basic resummed result with full NNLL resummation matched to NLO predictions from MCFM for the $\phi^*$ distribution 
normalised to have unit area\footnote{We have checked that this normalisation differs from the NLO inclusive rate obtained with MCFM by $\ord\left(\as^2 \right)$ terms, as one would expect.} along with a comparison to pure fixed-order predictions from MCFM, with the $\phi^*$ range being that over which the D\O\ collaboration has collected data \cite{D0phi}. The curves have been obtained for $p \bar{p}$ collisions at $1.96$~TeV with the CTEQ6m set of parton densities~\cite{cteq6m}, with the value of the strong coupling taken from the fit, $\as(M_Z)=0.1179$.
We consider leptons with invariant mass  $70$~GeV$<M<110$~GeV, transverse momenta $l_{T1,2}>15$~GeV and pseudorapidities $|\eta_{1,2}| < 2$ in the central rapidity bin of the vector boson $|y|<1$. This corresponds to the set of cuts adopted by \d0 for muons. It is noticeable that the matched result and the fixed-order predictions agree at large $\phi^*$ while the matched result behaves exactly like the resummation at smaller $\phi^*$ and the pure fixed-order result is seen to grow significantly at low $\phi^*$ reflecting the logarithmic divergences contained therein.  We have checked that the inclusion of full NNLL accuracy in $R(b)$ has a negligible impact on our previous predictions for the same quantity,  where we included only the first term of the NNLL function $g^{(3)}$~\cite{BDMphi}.

We note that a difference from the $Q_T$ distribution is the absence of a peak in our final resummed result. As we have observed before this is related to the fact that two distinct mechanisms can produce a low $\phi^*$ value: inhibition of gluon radiation which results in a Sudakov form factor 
and the vectorial cancellation between emissions that contribute to the one-dimensional sum in Eq.~\eqref{eq:oned}. The latter mechanism which corresponds to the small-$b$ region of the $b$ integral dominates the Sudakov behaviour at 
small values of $\phi^*$ so as to wash out the Sudakov peak in the $\phi^*$ distribution which simply rises to a constant value. 

Having presented and discussed our basic resummed result and its matching to fixed-order we now turn to the question of assessing the uncertainty on the matched resummed result, as manifested by a dependence on the various arbitrary 
perturbative scales contained in the full result, which in the discussion above had all been set equal to the pair invariant mass $M$. The study of perturbative uncertainty is an important step as, while one should observe a significantly smaller scale uncertainty at low $\phi^*$ for NNLL resummation as opposed to NLL resummation, one would nevertheless expect it to be important in the eventual comparisons to data and would certainly wish to take it into account before reaching conclusions on the size of non-perturbative effects.

\subsection{Perturbative uncertainties}\label{sec:PTunc}
\begin{figure}
\begin{center}
\includegraphics[width=0.49 \textwidth]{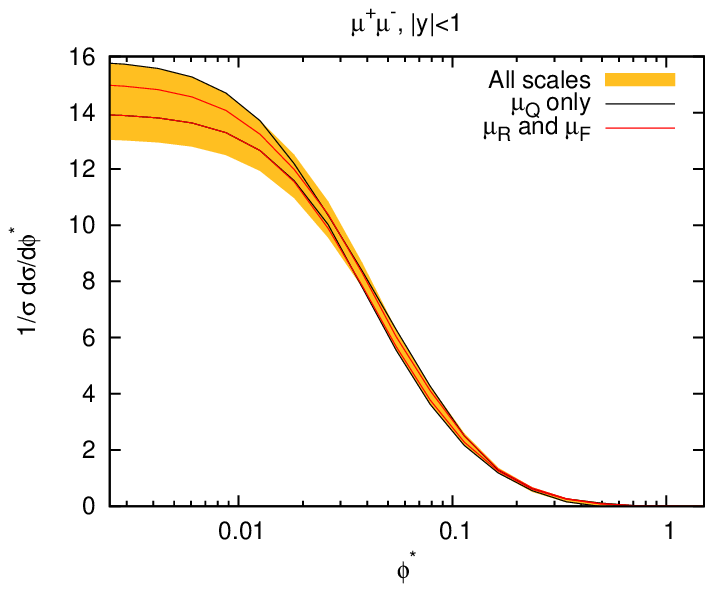}
\includegraphics[width=0.49 \textwidth]{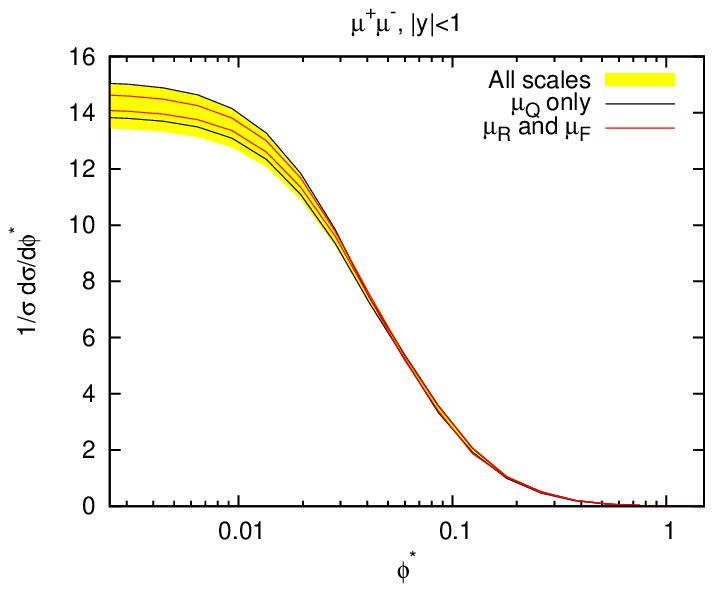}
\includegraphics[width=0.7 \textwidth]{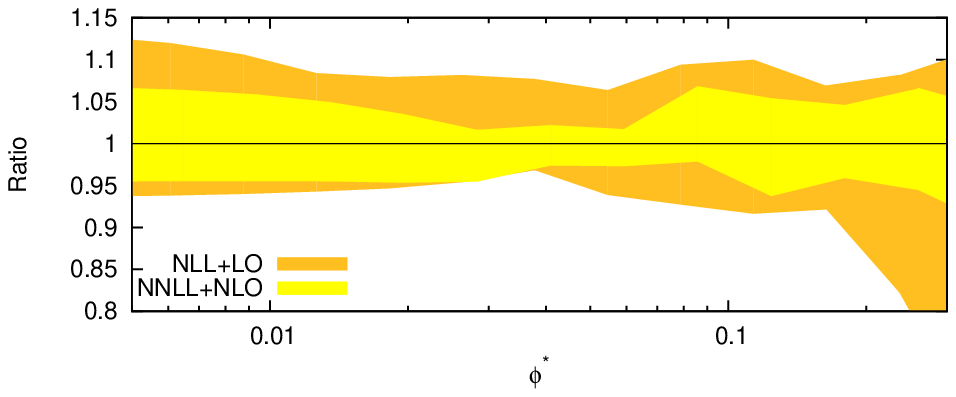}
\caption{Study of the perturbative uncertainties in case of NLL+LO (on the top left) and NNLL+NLO (on the top right). The black lines represent the variation of the resummation scale ${M}/{2}\le\mu_Q\le  2M$, while the red ones the variation of renormalisation and factorisation scales ${M}/{2}\le\{\mu_R, \mu_F\}\le  2M$, with ${1}/{2}\le\mu_R/ \mu_F\le  2$.  The orange (yellow) band is obtained by varying all scales independently, requiring the ratio of any two of them to be between $1/2$  and $2$.
At the bottom the uncertainty bands are normalised to the curve with all scales set to the pair mass}\label{fig:unc}
\end{center}
\end{figure}
Here we shall deal with the issue of the perturbative uncertainty afflicting our calculation. 
In general, one can consider the strong coupling to be evaluated, as usual, at some renormalisation scale $\mu_R$ which, although of the order of $M$, is not the same. Very similarly, the factorisation scale $\mu_F$, which sets the scale where the pdfs are evolved from, ought not to be exactly equal to $M$. Another arbitrary scale in our perturbative calculation enters the argument of the logarithms we are resumming, $\ln(\bar{b}^2M^2)\to \ln(\bar{b}^2\mu_Q^2)$, where we refer to $\mu_Q$ as the resummation scale. For analogous studies for the $Q_T$ spectrum see, for instance, Ref.~\cite{FlorenceDY}. Keeping the full dependence on those scales the resummed expression in Eq.~(\ref{eq:resummed})  becomes
\bea
\label{eq:resummed_muQ}
\frac{ {\rm d} \sigma}{ {\rm d} \phi^*} \left( \phi^*,M, \cos \theta^*, y \right) &=& \frac{\pi \alpha^2}{s N_c} 
\int_0^{\infty} {\rm d} b\, M \,\cos \left(bM \phi^* \right) 
e^{-R(\bar{b},M, \mu_Q,\mu_R)} \nonumber \\ &&\times\,\Sigma \left(x_1,x_2,\cos\theta^*, b,M,\mu_Q,\mu_R,\mu_F \right)\,.
\eea
Explicit formulae are reported in Appendices~\ref{app:res} and~\ref{app:radiator}.
The resummation is then matched to a fixed-order calculation, which also depends on renormalisation and factorisation scales. 
The dependence of the resummed and matched result on these arbitrary scales is one order higher than the accuracy we are working at, i.e. it affects terms which are at least N$^3$LL and NNLO. Thus,  varying them around the dilepton invariant mass $M$ provides us with an estimate of the size of those perturbative contributions which are beyond our accuracy. In doing so our theoretical prediction becomes a band, and the \emph{central value}, where all the scales are set equal to each other and to the pair mass, does not have any special physical meaning.

In Fig.~\ref{fig:unc} we perform a study of the theoretical uncertainty that affects our result. As an example we consider the set of \do cuts for the muons in the central rapidity bin $|y|<1$, but similar results can be found by looking at the other bins. We study two different levels of accuracy: NLL+LO, on the left, and NNLL+NLO, on the right.   The black lines represent the variation of the resummation scale ${M}/{2}\le\mu_Q\le  2M$, while the red ones represent the variation of renormalisation and factorisation scales ${M}/{2}\le\{\mu_R, \mu_F\}\le  2M$, with ${1}/{2}\le\mu_R/ \mu_F\le  2$.  The orange (yellow) band is obtained by varying all scales independently, requiring the ratio of any two of them to be between $1/2$  and $2$.
In the same figure, at the bottom, the uncertainty bands are normalised to the curve with all scales set to the pair mass.
 We note that at low $\phi^*$ the dominant source of uncertainty is given by the $\mu_Q$ variation, i.e. it comes from sub-leading logarithmic terms in the resummation. We also note that the uncertainty in the small $\phi^*$ region is almost halved in going from NLL+LO to NNLL+NLO. In fact we have that the size of the band is $\ord\left(20 \%\right)$ in the former case, while it is  $\ord\left(10 \%\right)$ in the latter. This is consistent with the NNLL+NLO uncertainty band found in the case of the $Q_T$ spectrum~\cite{FlorenceDY}.

Another source of theoretical uncertainty comes from the parton distribution functions. However, we find that these effects mostly cancel once the distribution has been normalised to the inclusive rate, leaving an uncertainty at the percent level, much smaller than the band obtained with scale variations.

Having estimated the theoretical uncertainty of our calculation, we can now compare it to the experimental data. This is the topic of the next section.

\section{Comparisons to data}
\begin{figure}
\begin{center}
\includegraphics[width=0.49\textwidth]{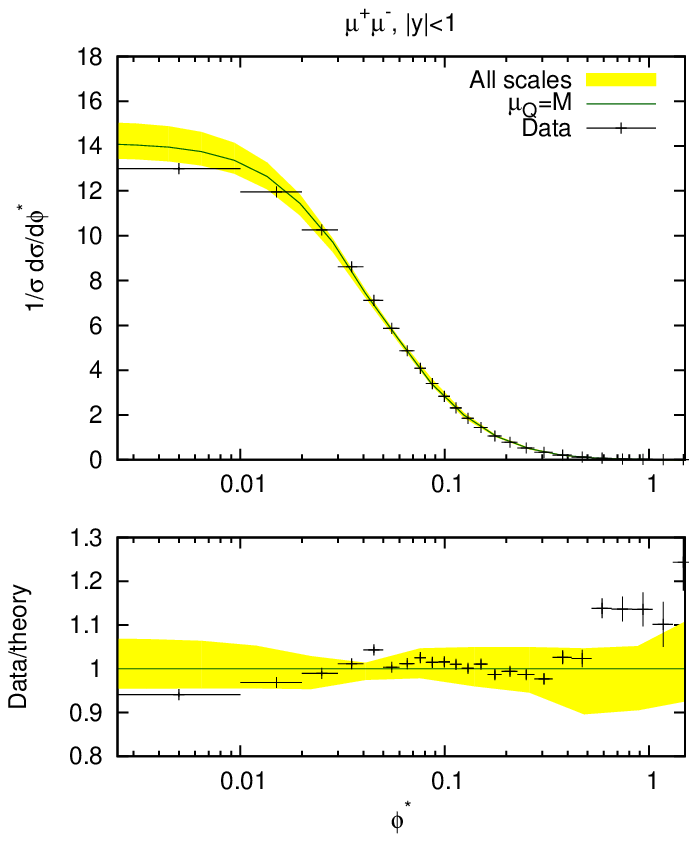}
\includegraphics[width=0.49\textwidth]{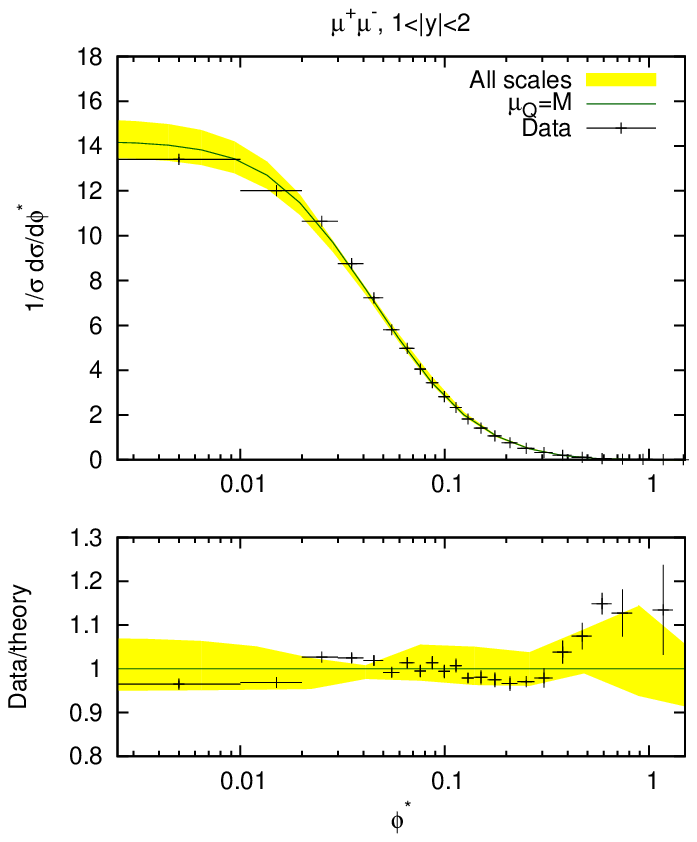}
\caption{Comparison of the theoretical prediction NNLL+NLO for the $\phi^*$ distribution to the experimental data collected by the  \d0 collaboration in the case of muons, in two different vector boson rapidity bins, $|y|<$1, on the left and $1<|y|<2$ on the right. }\label{fig:muon}
\end{center}
\end{figure}

\begin{figure}
\begin{center}
\includegraphics[width=0.49\textwidth]{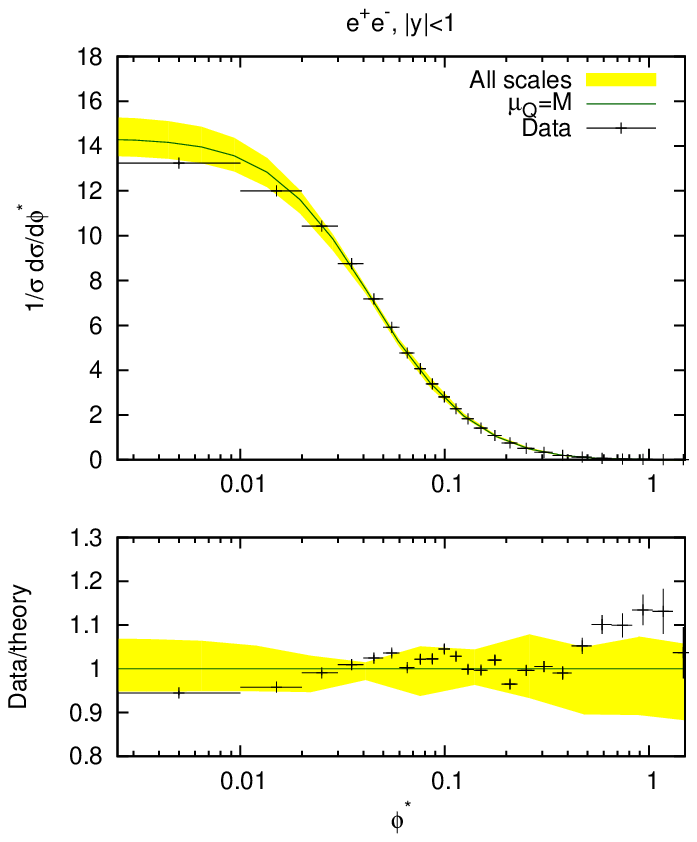}
\includegraphics[width=0.49\textwidth]{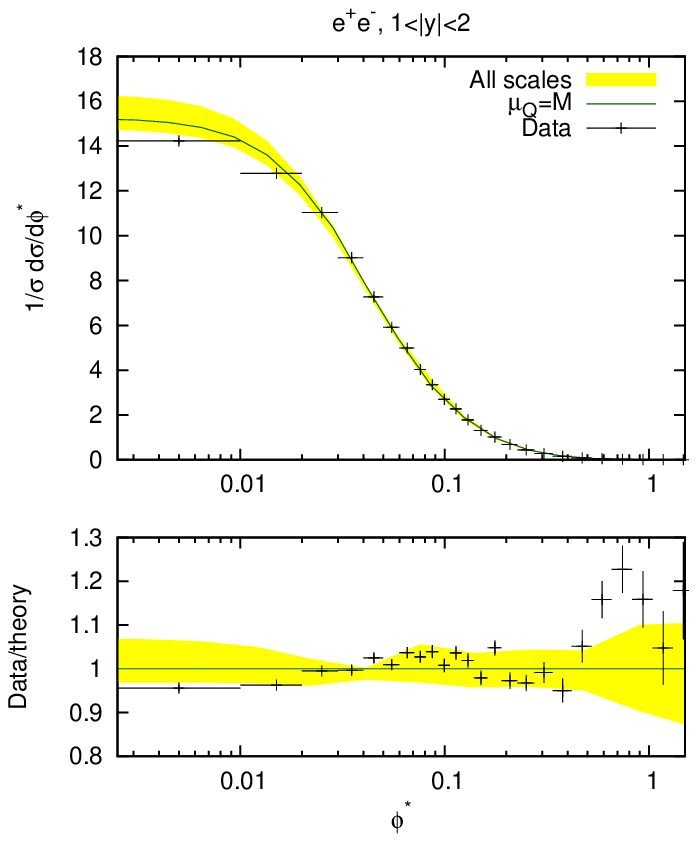}
\includegraphics[width=0.49\textwidth]{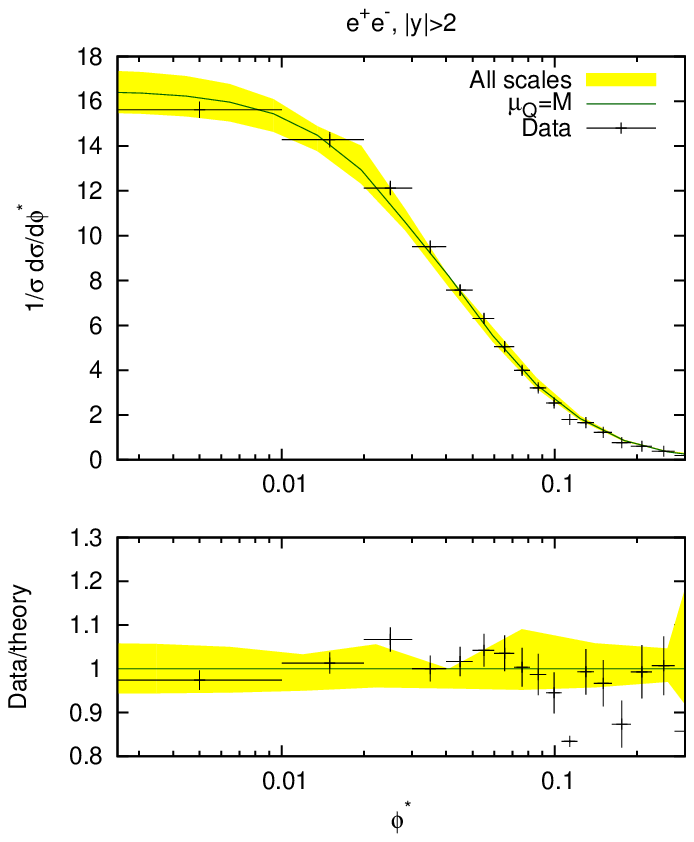}
\caption{Comparison of the theoretical prediction NNLL+NLO for the $\phi^*$ distribution to the experimental data collected by the  \d0 collaboration in the case of electrons, in three different vector boson rapidity bins, $|y|<$1, on the top left, $1<|y|<2$ on the top right and $|y|>2$ at the bottom.}\label{fig:elec}
\end{center}
\end{figure}

The \d0 experiment at the Fermilab Tevatron performed a measurement of the $\phi^*$ distribution in bins of $|y|$, using 7.3 fb$^{-1}$ of data~\cite{D0phi}. Different kinematical cuts are applied for muons and electrons.
More specifically, the invariant mass of the leptons must lie in the range $70$~GeV$<M<110$~GeV.  Moreover, muons must satisfy $l_{T1,2}>15$~GeV and $|\eta_{1,2}| < 2$, while electrons must satisfy $l_{T1,2}>20$~GeV and $|\eta_{1,2}| < 1.1$ or $1.5< |\eta_{1,2}| < 3$. 
We now compare our theoretical calculations to the data, taking fully into account these experimental cuts.  The theoretical predictions are normalised to the area under the curve, as  are the data.  
In Fig.~\ref{fig:muon} we compare the data to our theoretical prediction for the muon channel, in the central rapidity bin $|y|<1$ (on the left) and in the outer one $1<|y|<2$ (on the right). The yellow bands represent the theoretical uncertainty obtained  by varying the three scales $\mu_Q$, $\mu_R$ and $\mu_F$ independently, as explained in Sec.~\ref{sec:PTunc}. The solid (green) curve is the theoretical prediction obtained with all scales set equal to the pair mass $\mu_Q=\mu_R=\mu_F=M$. The plots in Fig.~\ref{fig:elec} instead show the comparison to the electron data. In this case we have three different rapidity bins: $|y|<1$ (on the top left), $1<|y|<2$ (on the top right) and $|y|>2$ (at the bottom). The last rapidity bin for the electrons provides an opportunity to probe fairly small-$x$ values ($x < 10^{-2}$) with conventional resummation techniques. In this study we are mostly interested in the low $\phi^*$ region of the distribution, hence we do not show the large $\phi^*$ tail of the outer rapidity bin because the experimental errors, as well as the theoretical uncertainties, are rather large and the comparison is not very instructive.  

 We observe, on the whole, within the scale uncertainties, an excellent agreement over a large range of $\phi^*$ for both muons and electrons in all rapidity bins. This includes the lowest $\phi^*$ values where one may expect non-perturbative effects to play some role. 
 
\begin{figure}
\begin{center}
\includegraphics[width=0.49\textwidth]{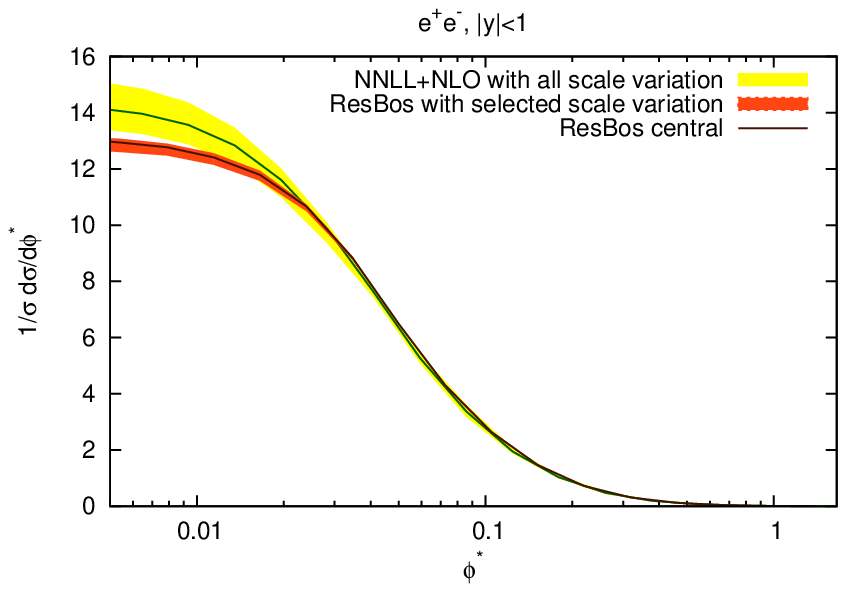}
\includegraphics[width=0.49\textwidth]{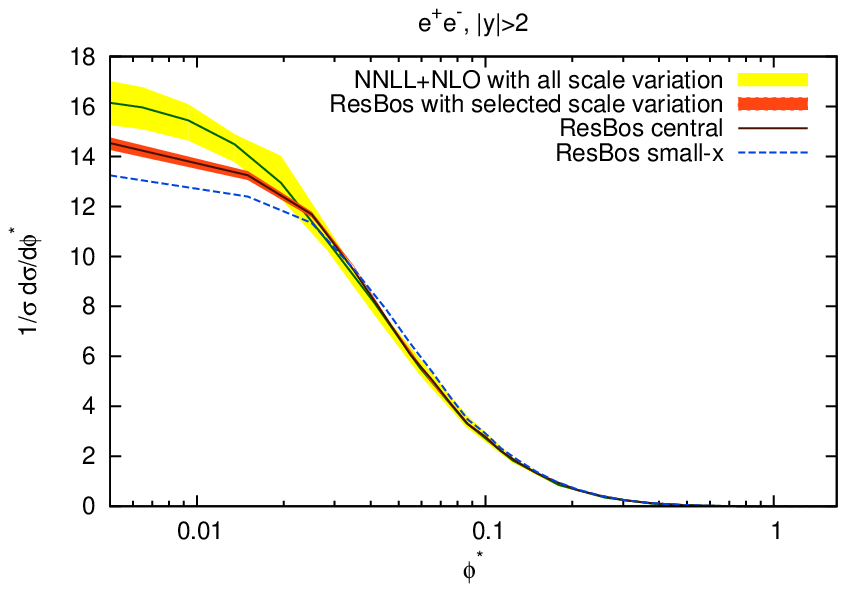}
\caption{Comparison of our NNLL+NLO theoretical prediction for the $\phi^*$ distribution to the one obtained with the program RESBOS, in the case of electrons in the central rapidity bin $|y|<1$, on the left, and the forward one $|y|>2$, on the right.}\label{fig:resbos}
\end{center}
\end{figure}
It is worth noting that we obtain a comparable description of \d0 data to that provided by the commonly used event generator RESBOS~\cite{resbos1, resbos2}, however, with different underlying physical assumptions. This is shown in Fig.~\ref{fig:resbos}, where we compare our NNLL+NLO theoretical predictions, in case of electrons in the central (left) and forward (right) rapidity bins, to the ones obtained with the RESBOS, run with the same event selection as in Ref.~\cite{D0phi}. We note that the height of the plateau predicted by RESBOS is lower than ours, because of the presence of non-perturbative effects, modelled by the BLNY form factor\footnote{The BLNY form factor~\cite{BLNY} is primarily controlled by one parameter $g_2$. We use the default value $g_2 = 0.68$~GeV$^2$.}. 
There is a striking difference between the sizes of the theoretical uncertainties of the two predictions. Following~\cite{D0phi}, the RESBOS uncertainty is obtained by varying $\mu_R$ and $\mu_F$ simultaneously by a factor of two, although we do not include here PDFs uncertainties. This procedure does not capture the main source of uncertainty, which, in our approach, is given by variations of the resummation scale $\mu_Q$.
Finally, in the forward rapidity bin, we also we show the RESBOS prediction including small-$x$ broadening effects which were obtained by fitting RESBOS to semi-inclusive-deep-inelastic-scattering data~\cite{QTsmallx}. The resulting distribution is broader and, consequently, its height is lower. Small-$x$ broadening is disfavoured by the data, as discussed in~\cite{D0phi}.

\begin{figure}
\begin{center}
\includegraphics[width=0.49\textwidth]{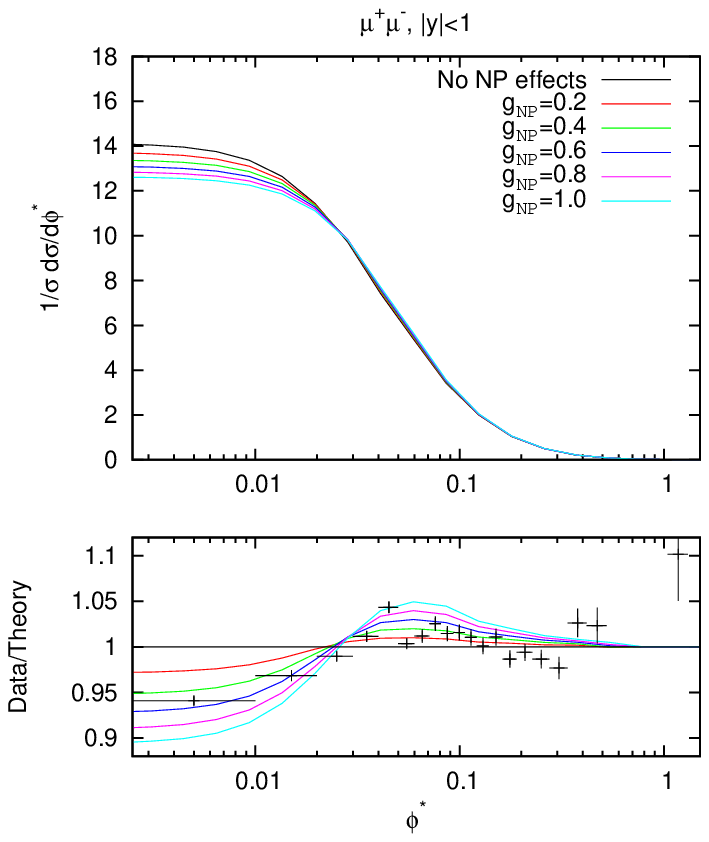}
\includegraphics[width=0.49\textwidth]{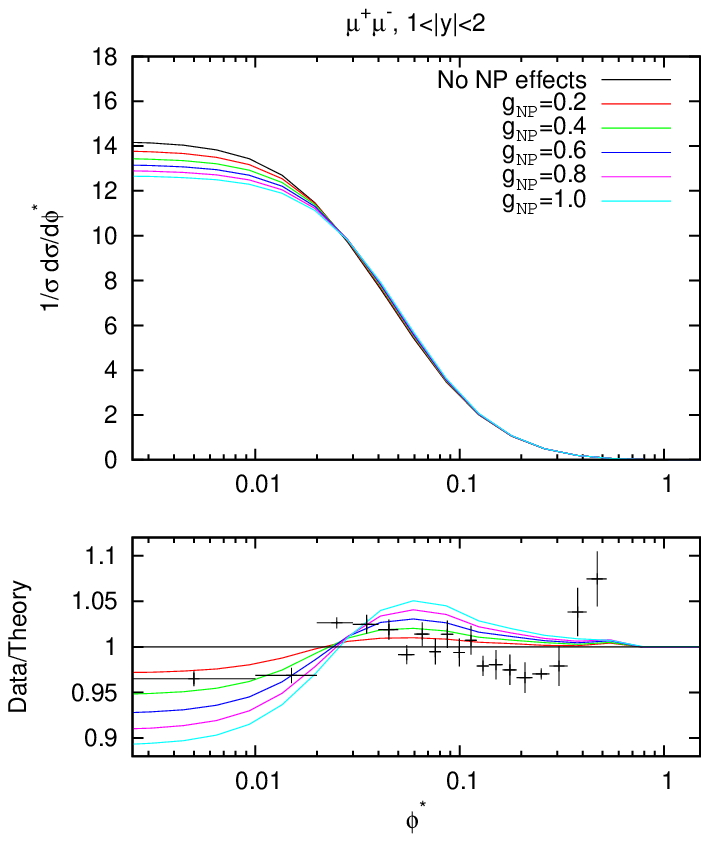}
\caption{Distributions obtained with different values of the non-perturbative parameter $g_{\rm NP}$, in the different rapidity bins for muons. All curves are obtained setting $\mu_Q=\mu_R=\mu_F=M$.}\label{fig:NPm}
\end{center}
\end{figure}

\begin{figure}
\begin{center}
\includegraphics[width=0.49\textwidth]{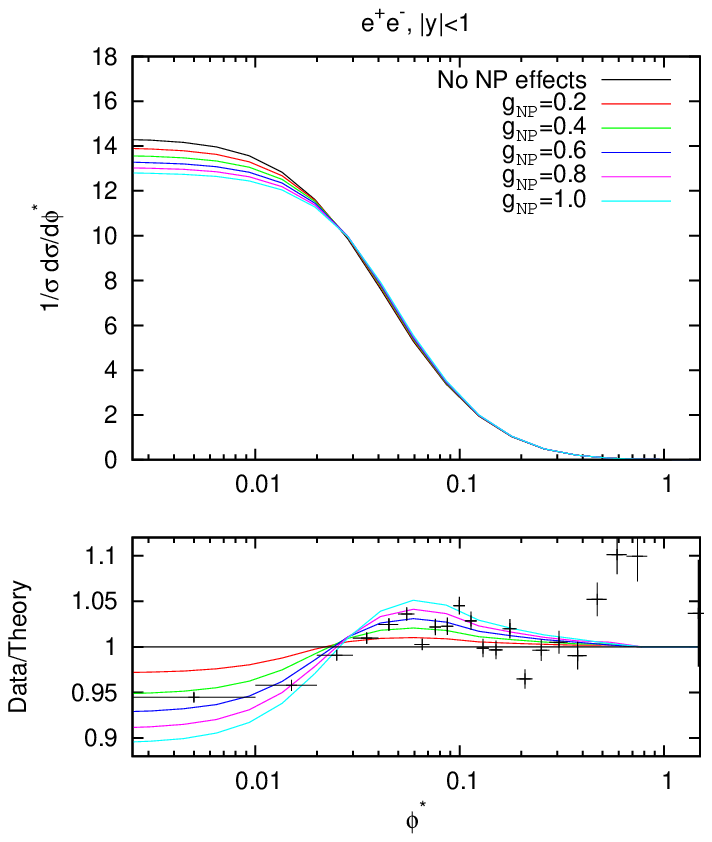}
\includegraphics[width=0.49\textwidth]{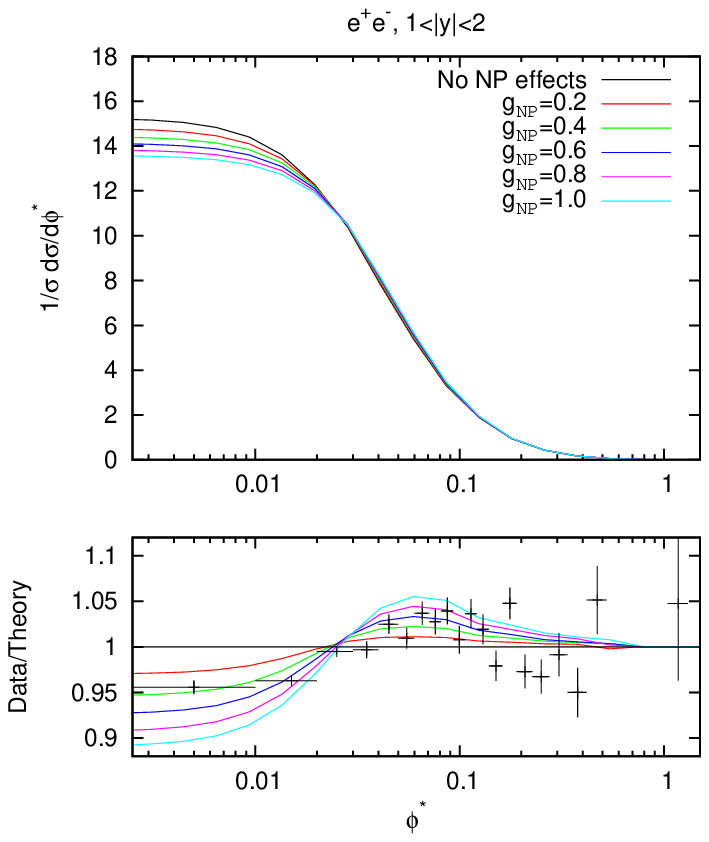}
\includegraphics[width=0.49\textwidth]{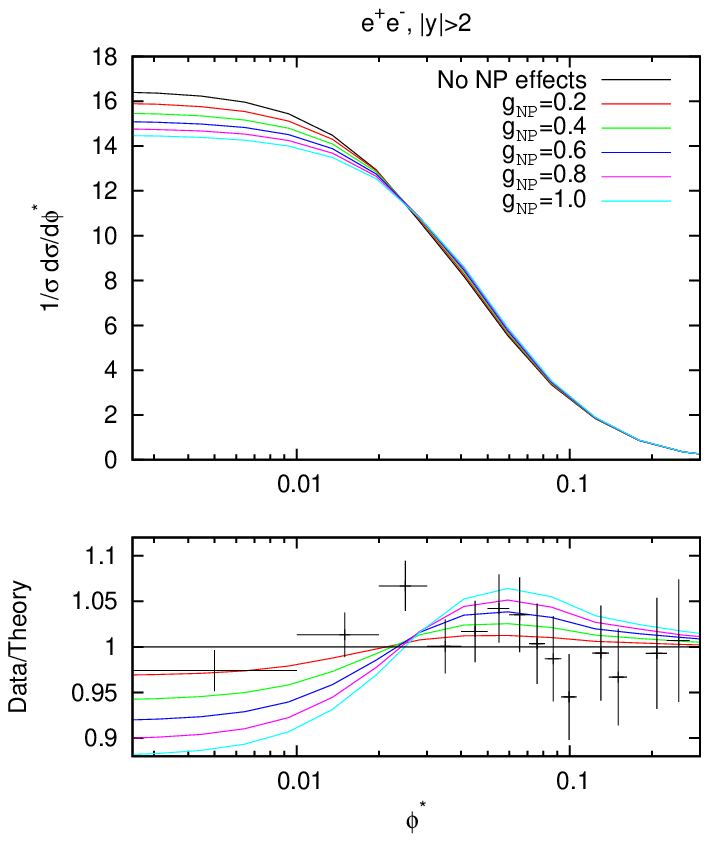}
\caption{Distributions obtained with different values of the non-perturbative parameter $g_{\rm NP}$, in the different rapidity bins for electrons. All curves are obtained setting $\mu_Q=\mu_R=\mu_F=M$.}\label{fig:NPe}
\end{center}
\end{figure}

Our analysis of the theoretical uncertainty has led us to the conclusion that it is hard to make any statements on the size of non-perturbative effects, except to set a range of values based on the upper and lower edges of our uncertainty band.
Nevertheless, it is possible to take our central values, with all scales set equal to the pair mass, and correct it to the data by adding a non-perturbative smearing to the radiator Eq.~(\ref{radiator}):
\beq\label{nprad}
R_{\rm NP}\left(\bar{b} M\right) = R\left(\bar{b}M\right) + g_{\rm NP} b^2\,,
\eeq
 which corresponds to assigning a Gaussian-smeared intrinsic transverse momentum to the incoming partons. We can now look for the NP parameter $g_{\rm NP}$ which gives the best description of the data. The results are shown in Fig.~\ref{fig:NPm} and~\ref{fig:NPe}, with different choices of $g_{\rm NP}$, inspired by the literature on $Q_T$ resummation.
 We note that the spread of results obtained by varying the NP parameter from $g_{\rm NP}=0$~GeV$^{2}$ to $g_{\rm NP}=1$~GeV$^{2}$ is similar to the band describing the perturbative uncertainty.
  The best value of the NP parameter is different in different bins. We obtain $g_{\rm NP}\simeq 0.5$~GeV$^{2}$ for central rapidities and $g_{\rm NP}\simeq 0.3$~GeV$^{2}$ in the more forward regions. 
However, we stress that this procedure is misleading since we have no reason to assign any special role to the curve where all the perturbative scales are set equal to the dilepton mass. In doing so we would be ascribing the perturbative uncertainty to a universal NP parameter and the use of this parameter for studies of related variables could result in erroneous conclusions. 

\section{Comments and conclusions}

In the current paper we have carried out a complete phenomenological study for the $\phi^*$ variable recently measured by the D\O\ collaboration \cite{D0phi}. We have included resummation to the next-to--next-to-leading logarithmic level and matched the resummation to the full NLO calculation from MCFM \cite{mcfm}. We carried out a study estimating the theoretical uncertainty on the matched resummed prediction and provided comparisons to data. We also investigated the potential role of non-perturbative effects in the comparison of theory to experiment.

One of the aims of our study was to exploit the fact that the measurement errors on $\phi^*$ are significantly smaller than for the standard $Q_T$ variable 
\cite{WV}, to study as precisely as possible the role and accuracy of 
resummed calculations and eventually also non-perturbative effects. In this context, we noted that the pure fixed-order (NLO) estimate was inadequate to describe the $\phi^*$ distribution over a very wide range of measured $\phi^*$ values, indicating both the need for resummation as well an an opportunity to compare different resummation approaches (including parton shower event generators) with a view to identifying significant physics differences and further developing resummation tools and event generator models. 

We observed that our basic matched resummed results depend as usual on various arbitrary scales, specifically, the renormalisation scale $\mu_R$, the factorisation scale $\mu_F$ and the resummation scale $\mu_Q$. Varying these scales gives an idea of the impact of sub-leading terms omitted from our treatment and, as a result, we derived an uncertainty band for our prediction, which reflects the theoretical error on our estimate.

 We found that, within the aforementioned error band, we get an excellent description of the $\phi^*$ distribution with hardly any variation in the quality of our description in the different bins in $y$, the lepton pair rapidity. We would thus deduce that the size of non-perturbative effects cannot be ascertained unless one reduces further the theoretical uncertainty with, for instance, an ${\mathrm{N^3LL}}$ resummation. We also directly studied non-perturbative effects and found that reasonable values of a non-perturbative Gaussian smearing generated a spread of results comparable in size to our perturbative uncertainty. We stress that taking our central value and correcting it to the data with a non-perturbative effect, while possible, is misleading since it leads potentially to ascribing physics, which could be of perturbative origin, to universal non-perturbative effects. A preferable approach to non-perturbative effects from our viewpoint would be to theoretically study them for the $\phi^*$ variable using for instance the techniques based on a universal infrared finite extension of $\alpha_s$ as applied for instance to $Q_T$ resummation \cite{SmyGuff} and the related case of energy-energy correlations in $e^{+}e^{-}$ annihilation \cite{DokWebMar}. We postpone this investigation to future work. 

As far as further developments are concerned, one could envisage, given the excellent agreement of the precise perturbative results with the $\phi^*$ data, that with a suitable fit range one may use the results presented here to fit for 
the strong coupling, $\alpha_s$, from the D\O\ data. On the more theoretical 
side one useful development would be to try and formulate the resummation of the $\phi^*$ variable in a $b$-space-independent framework. The main reason for this is that the use of $b$-space requires us to provide arbitrary prescriptions to deal with regions of integration which are not strictly under the control 
of resummation, as well as serves to potentially contaminate our results with non-perturbative effects via the use of parameters such as $b_{\mathrm{max}}$, designed to avoid the Landau pole. While we have noted the stability of our results against reasonable variations of such parameters, it is clearly desirable to remove them altogether, if possible.
Resummed studies exist for the $Q_T$ distribution which do not employ the $b$-space formalism, both within traditional resummation methods \cite{EV} and within the techniques of soft-collinear effective theory \cite{BecNeu,BecNeu2} and it would be interesting to see if such methods could be used more generally for variables such as $\phi^*$ and other similar studies. We note that similar methods to those used in our current study have also been used to study azimuthal decorrelations between coloured final state particles (specifically for jets see Ref.~\cite{BanDasDel08}), which represents an extension of the $Q_T$ resummation formalism to processes with colour in the final state.

Lastly, since it is straightforward to extend our studies to the LHC, we would also advocate measurement of the $\phi^*$ variable at the LHC 
and suggest that studying $\phi^*$ together with the $Q_T$ distribution \cite{AtlasZpT} would provide invaluable information especially when detailed comparisons are made to the variety of theoretical calculations and Monte Carlo models that are currently available, including our forthcoming results.

  {\bf Acknowledgments}
We thank Terry Wyatt and Mika Vesterinen for many useful discussions. 
Two of us (AB) and (MD) gratefully acknowledge the hospitality and support of the Galileo Galilei Institute (GGI), Firenze, during the completion of this work. One of us (AB) also acknowledges the hospitality of the Manchester Particle Physics  group while carrying out some of this work.
This work is supported by UK's STFC and by the Swiss National Science Foundation under
contract SNF 200020-126632.

\appendix
\section{Resummed result}\label{app:res}
The resummed expression for the $\phi^*$ distribution has been reported in Eq.~(\ref{eq:resummed}), with all the perturbative scales set equal to the dilepton mass and in Eq.~(\ref{eq:resummed_muQ}), where they were kept separate.
The function $\Sigma$ is given by
\bea
\Sigma \left(x_1,x_2,\cos\theta^*, b,M,\mu_Q,\mu_R,\mu_F \right)&= &e^{-\frac{\as(\mu_R)}{\pi} g^{\rm corr}\left(\frac{M}{\mu_Q}\right) } \times
\nonumber \\ &&\widetilde{\Sigma} \left(x_1,x_2,\cos\theta^*, b,M,\mu_Q,\mu_R,\mu_F \right),
\eea
where
\beq
 g^{\rm corr}\left(\frac{M}{\mu_Q}\right) = \frac{C_F}{2} \ln^2\frac{M^2}{\mu_Q^2}-\frac{3}{2}C_F \ln \frac{M^2}{\mu_Q^2}.
 \eeq
This contribution ensures full control of the logarithms at NNLL in the expansion,  that is the dependence on the resummation scale $\mu_Q$ starts at N$^3$LL.
Note that $\Sigma$ acquires a dependence on the impact parameter $b$ because of the resummation of logarithms of $b$ via DGLAP evolution, which then determines the scale of the parton distribution functions embedded in $\widetilde{\Sigma}$.
Including the contributions from the $Z$ as well as from the virtual photon, we  have 
\begin{eqnarray} \label{sigma}
\widetilde{\Sigma} &=& (1+\cos^2 \theta^*)\left(Q_q^2- 2 Q_q V_l V_q \chi_1 +(A_l^2+V_l^2)(A_q^2+V_q^2) \chi_2\right)  {\mathcal{F}}_q^+
 \nonumber \\  &&+  \cos \theta^* (-4 Q_q A_l A_q \chi_1+ 8 A_l V_l A_q V_q \chi_2)  {\mathcal{F}}_q^{-},
\end{eqnarray}
where a sum over quark flavours $q$ is implied. The above equation is naturally written as the sum of the two terms with different angular dependence: the first one is proportional to $(1+\cos^2 \theta^*)$ and represents the parity conserving piece of the electro-weak interaction, while the term involving $\cos \theta^*$ is the parity violating piece. We notice that upon integration over the full $\theta^*$ range, as well as over symmetric intervals, the parity violating term vanishes.
The coefficients $A_{l,q}$ and $V_{l,q}$ are the electroweak couplings for lepton $l$ and parton $q$, explicitly given by:
\begin{equation}
A_f = T^3_f \quad {\rm  and } \quad V_f = T^3_f- 2 Q_f \sin^2 \theta_W, \quad f= l,q \,,
\end{equation}
where $T^3_f$ is the third component of the isospin.
 We also have introduced
\begin{eqnarray}
\chi_1 &=& \kappa \frac{M^2 \left(M^2-M_Z^2 \right)}{(M^2-M_Z^2)^2 +\Gamma_Z^2 M_Z^2}, \nonumber \\
\chi_2 &=& \kappa^2 \frac{M^4}{(M^2-M_Z^2)^2 +\Gamma_Z^2 M_Z^2},  \nonumber \\
\kappa  &=& \frac{\sqrt{2} G_F M_Z^2}{4 \pi \alpha}.
\end{eqnarray}

In Eq.~(\ref{sigma}), $\mathcal{F}^{\pm}$ are explicitly given by
\bea
\label{eq:Fdef}
{\mathcal{F}}_q^{\pm}&=&
 \left(\bf{C} \otimes \bf{f}_1\right)_q(x_1,\bar{b},\mu_Q,\mu_R,\mu_F)\left(\bf{C} \otimes \bf{f}_2\right)_{\bar{q}}(x_2,\bar{b},\mu_Q,\mu_R,\mu_F) 
 \nonumber\\ &\pm& \left(\bf{C} \otimes \bf{f}_1\right)_{\bar{q}}(x_1,\bar{b},\mu_Q,\mu_R,\mu_F)\left(\bf{C} \otimes \bf{f}_2\right)_{q}(x_2,\bar{b},\mu_Q,\mu_R,\mu_F).
\eea

The convolutions involving the matrix of coefficient functions $\bf{C}$ and 
the vector of parton densities $\bf{f}_{1,2}$ for incoming hadrons $1$ and $2$ respectively can be explicitly written as 
\begin{equation}
\label{eq:conv}
\left(\bf{C} \otimes \bf{f}_i\right)_q(x_i,\bar{b},\mu_Q,\mu_R,\mu_F)= \int_{x_i}^{1} \frac{dz}{z} C_{q \alpha} \left (\alpha_s \left( \frac{\mu_R}{\bar{b}\mu_Q} \right),\frac{x_i}{z}, \frac{\mu_F}{\mu_Q}\right)
f_i^\alpha\left(z,\frac{\mu_F}{\bar{b}\mu_Q}\right),
\end{equation}
where $i=1,2$ and a sum over all flavours $\alpha$ is implied. The combinations of scales in the argument of the strong coupling and of the pdfs are such that, when $\bar{b}=\bar{b}_{\mathrm{min}}=\mu_Q^{-1}$ and, consequently $R=0$,  they reduce to $\mu_R$ and $\mu_F$, respectively. 

The coefficient functions $C_{q \alpha}$ represent perturbative corrections to the 
collinear branching of an incoming parton $\alpha$ to a parton $q$ which annihilates with $\bar{q}$ to form the $Z$ boson. We note that the collinear enhanced terms generated by such a branching are incorporated to our accuracy into the scale of the pdfs $f_i$ via their dependence on the impact parameter $b$. Thus the coefficient functions represent only the non-logarithmic constant terms. 
Because of their collinear origin, the scale of the coupling constant in the coefficient function is naturally of the same order as the one for the scale which pdfs are evolved to. However, there is in principle a hard (process-dependent) contribution as well, coming from virtual corrections, characterised by $k_T\sim M$. We consistently choose to work in the so-called ``Drell-Yan scheme'' in which one sets this contribution to one~\cite{CatGrazll}.

The coefficient function admits the following expansion in the strong coupling constant:
\begin{equation}
C_{q \alpha}\left (\alpha_s \left( \frac{\mu_R}{\bar{b}\mu_Q} \right),x, \frac{\mu_F}{\mu_Q}\right) = \delta_{q \alpha}\delta \left(1-x \right)+\frac{\alpha_s\left(\mu_R/(\bar{b}\mu_Q)\right)}{2 \pi}C^{(1)}_{q \alpha}\left (x,\frac{\mu_F}{\mu_Q} \right ) +\mathcal{O} \left(\alpha_s^2 \right).
\end{equation}
In this work, we consistently resum next-to--next-to-leading logarithms in the radiator. When the exponential is expanded, the $\mathcal{O} \left(\alpha_s^2 \right)$ contribution to coefficient function generates terms which are of the same order as the ones in $g^{(3)}$  and hence, in principle, should also be included. However,  these contributions start at $\mathcal{O} \left(\alpha_s^3 \right)$ and hence we would expect their effect to be numerically as significant as  the one due to the coefficient $A^{(3)}$ , which we have found to be small. Therefore,  in the current paper we use the first order approximation of the coefficient functions and we leave the inclusion of the $\mathcal{O} \left(\alpha_s^2 \right)$ contributions to future work.
The explicit form of the $\mathcal{O} \left(\alpha_s\right)$ coefficient functions is
\bea
 C^{(1)}_{q\bar{q}}\left(x,\frac{\mu_F}{\mu_Q}\right) &= &  C_F  \left(\frac{\pi^2}{2}-4\right) \delta(1-x)+ (1-x)  + \ln \frac{\mu_Q^2}{\mu_F^2} P^{(0)}_{qq}(x)\,, \\
 C^{(1)}_{qg}\left(x,\frac{\mu_F}{\mu_Q}\right) &= &  x (1-x)+  \ln\frac{\mu_Q^2}{\mu_F^2}P^{(0)}_{qg}(x) \,,
\eea
where $P^{(0)}_{\alpha \beta}$ are the LO DGLAP splitting functions.

\section{The radiator to NNLL}\label{app:radiator}
In this appendix we collect the explicit expression for the functions $g_i$, which enter into the radiator:
\bea\label{radiator_muQ}
R\left(\bar{b}\mu_Q,\frac{M}{\mu_Q},\frac{\mu_Q}{\mu_R};\as(\mu_R)\right) &=& L g^{(1)}(\as L) + g^{(2)}\left(\as L, \frac{M}{\mu_Q},\frac{\mu_Q}{\mu_R}\right) \nonumber \\ &&+ \frac{\as}{\pi} g^{(3)}\left(\as L, \frac{M}{\mu_Q},\frac{\mu_Q}{\mu_R}\right)\,,
\eea
where $L=\ln(\bar{b}^2\mu_Q^2)$ and $\as = \as(\mu_R)$ is the $\overline{\rm MS}$ coupling. The explicit expressions for the functions $g^{(i)}$ are

\bea \label{gi}
  g^{(1)}(\lambda)& =&- \frac{A^{(1)}}{\pi\beta_0} \frac{\lambda+\ln{(1-\lambda)}}{\lambda},\\
  g^{(2)}({\lambda})&=&  -\frac{B^{(1)}}{\pi \beta_0} \ln(1-\lambda)
 +\frac{A^{(2)}} {\pi^2 \beta_0^2} \left (\frac{\lambda}{1-\lambda} +\ln(1-\lambda)\right)
 \nonumber \\
  &   -&\frac{A^{(1)} \beta_1}{\pi\beta_0^3}
  \left[ \frac{\lambda + \ln (1-\lambda)}{1-\lambda} + \frac{1}{2}
   \ln^2{(1-\lambda)} \right] -\frac{A^{(1)}}{\pi \beta_0} \left(\frac{\lambda}{1-\lambda}+\ln(1-\lambda) \right) \ln \frac{\mu_Q^2}{\mu_R^2} ,\nonumber \\ & &\\
  g^{(3)}(\lambda) &=& \frac{A^{(3)} } {2 \pi^2 \beta_0^2} \frac{\lambda^2}{(1-\lambda)^2}+ \frac{B^{(2)}}{\pi \beta_0}\frac{\lambda}{1-\lambda}
  -\frac{A^{(2)}\beta_1}{\pi \beta_0^3} \left( \frac{\lambda (3 \lambda-2)}{2 (1-\lambda)^2}-\frac{(1-2 \lambda) \ln (1-\lambda)}{(1-\lambda)^2}  \right) 
\nonumber  \\
  &-&\frac{A^{(1)}}{\beta_0^4} \left( \frac{\beta_1^2}{2} \frac{1- 2\lambda}{(1-\lambda)^2}\ln^2(1-\lambda)+ \frac{\ln(1-\lambda)}{1-\lambda} \left(\beta_0 \beta_2 (1-\lambda)+\beta_1^2 \lambda  \right) 
 \right.   \nonumber \\
  &+& \left. \frac{\lambda}{2 (1-\lambda)^2} \left(\beta_0 \beta_2 (2-3 \lambda)+\beta_1^2 \lambda \right)   \right)  \nonumber \\
 & -&\frac{B^{(1)}\beta_1}{\beta_0^2} \left(\frac{\lambda}{1-\lambda} +\ln(1-\lambda) \right)+\frac{A^{(1)}}{2}\frac{\lambda^2}{(1-\lambda)^2} \ln^2 \frac{\mu_Q^2}{\mu_R^2} \nonumber\\
 &-&\left( B^{(1)} \frac{\lambda}{1-\lambda}+ 
 \frac{A^{(2)}}{\pi \beta_0} \frac{\lambda^2}{(1-\lambda)^2} +A^{(1)} \frac{\beta_1}{\beta_0^2} \left(\frac{\lambda}{1-\lambda}+
 \frac{1-2 \lambda}{(1-\lambda)^2} \ln (1-\lambda) \right)   \right) \ln \frac{\mu_Q^2}{\mu_R^2},  \nonumber\\
\eea
with $\lambda =  \alpha_s(\mu_R)\,\beta_0 L$. 

The coefficients $A^{(i)}$ and $B^{(i)}$ are given by
\bea
A^{(1)} &=&  C_F \,, \\
A^{(2)} &=& \frac{C_F}{2} \left( C_A \left(\frac{67}{18} - \frac{\pi^2}{6} \right) - \frac{5}{9} n_f \right) \\
A^{(3)} &=&   \frac{C_F}{16} \left (C_A^2 \left(\frac{245}{6} - \frac{134}{27} \pi^2 + \frac{11}{45} \pi^4 + \frac{22}{3} \zeta_3 \right) + 
   \frac{1}{2} C_A n_f \left(-\frac{418}{27} + \frac{40}{27} \pi^2 - \frac{56}{3} \zeta_3 \right) \right. \\ \nonumber &&+ \left.
      \frac{1}{2}C_F n_f \left(-\frac{55}{3} + 16 \zeta_3 \right) - \frac{4}{27} n_f^2 \right) + \frac{1}{8} \pi \beta_0 d_2\,,\\
B^{(1)} &=& -\frac{3}{2} C_F + A^{(1)}\ln \frac{M^2}{\mu_Q^2}\,,\\
B^{(2)} &=& \frac{1}{4} \left( C_F^2\left(\pi^2 -\frac{3}{4} - 12 \zeta_3 \right) + 
      C_F C_A \left(\frac{11}{9} \pi^2 - \frac{193}{12} + 6 \zeta_3 \right) + 
      \frac{1}{2}C_F n_f \left(-\frac{4}{9} \pi^2 + \frac{17}{3} \right) \right)  \nonumber \\ &+&A^{(2)}\ln \frac{M^2}{\mu_Q^2}\,.
\eea
The coefficient $d_2$ has been recently determined~\cite{BecNeu}:
\beq
d_2=  C_F C_A \left ( \frac{808}{27} - 28 \zeta_3 \right) - \frac{112}{27} C_F n_f\,.
\eeq

We have also introduced the coefficients of the QCD $\beta$-function
\bea
\beta&=&- \as \left( \beta_0 +\beta_1 \as+\beta_2 \as^2 + \ord\left(\as^3 \right) \right), \nonumber \\
\beta_0 &=& \frac{11 C_A-2 n_f}{12 \pi}, \nonumber \\ \beta_1 &=&\frac{17 C_A^2-5C_A n_f -3 C_F n_f}{24 \pi^2},
\nonumber\\ \beta_2&=& \frac{  \frac{2857}{54} C_A^3 - \frac{1415}{54} C_A^2 n_f - \frac{205}{18} C_A C_F n_f + 
    C_F^2 n_f + \frac{79}{54} C_A n_f^2 + \frac{11}{9} C_FF n_f^2}{64 \pi^3}.
\eea

\end{document}